\documentclass[aps,pre,amsmath,amssymb,twocolumn]{revtex4}

\usepackage{graphicx}
\usepackage{amsmath}
\usepackage{verbatim}
\usepackage{xcolor} 

\usepackage{algorithmic}
\usepackage{algorithm}

\newcommand{\iu}{\mathrm{i}} 
\DeclareMathOperator{\sech}{sech}
\DeclareMathOperator{\Op}{Op}

\begin{document}
\title{Manakov system on metric graphs:\\ Modeling the reflectionless propagation of vector solitons in networks}
\author{J.R.~Yusupov$^{1,2}$, Kh.Sh.~Matyokubov$^{3}$, M.~Ehrhardt$^{4}$ and D.U.~Matrasulov$^{5}$}
\affiliation{$^1$Yeoju Technical Institute in Tashkent, 156 Usman Nasyr Str., 100121, Tashkent, Uzbekistan\\
$^2$National University of Uzbekistan, 2 Universitet Str., 100095, Tashkent, Uzbekistan\\
$^3$Urgench State University, 14 H. Olimjon Str., 220100 Urgench, Uzbekistan\\
$^4$Bergische Universit\"at Wuppertal, Gau{\ss}strasse 20, D-42119 Wuppertal, Germany\\
$^5$Turin Polytechnic University in Tashkent, 17 Niyazov Str., 100095, Tashkent, Uzbekistan}

\begin{abstract}
We consider the reflectionless transport of Manakov solitons in networks. 
The system is modelled in terms of the Manakov system on metric graphs subject to transparent boundary conditions at the branching points. 
Simple constraints combining the equivalent usual Kirchhoff vertex conditions with the transparent conditions are derived in terms of nonlinearity coefficients.
 Although the method is used for a metric star graph, an extension to more complicated graph topologies is easily possible.
\end{abstract}

\maketitle

\section{Introduction}
The Manakov system is an integrable system of coupled nonlinear Schr\"odinger equations (NLSE) that allows various soliton solutions. 
It is widely used in modelling vector solitons propagating in Kerr media (nonlinear optics), in describing matter waves (physics of Bose-Einstein condensation) and in describing the propagation of matter waves, in the description of the propagation of matter waves (physics of Bose-Einstein condensates (BEC)), in the propagation of orthogonally polarised beams in planar $AlGaAs$ waveguides \cite{Kang}, 
in ultrafast soliton switching devices \cite{Yang99} and in the modelling of rogue waves \cite{Zhong}. 
Logic gates and computational operations based on colliding Manakov vector solitons were studied in \cite{Steiglitz}. 
The dynamics of Manakov vector solitons in optical fibres was studied in Refs.\ \cite{Kivshar1, Agarwal1}. 
Various mathematical aspects of the Manakov system and properties of Manakov vector solitons were studied in \cite{Lakshmanan1, Lakshmanan2, Ablowitz1, Lakshmanan3, Kivshar2, Lakshmanan4, Feng, Ohta, Panos1, Frantzeskakis}.

In optical and opto-electronic applications, vector solitons are used as signal carriers, with signal transfer usually taking place in branched fibres and networks. 
For optimal functioning of such devices, signal losses must be avoided or minimised by achieving a minimum of soliton backscattering, i.e.\ the solitons are propagated without reflections. 
The successful solution of such a problem requires the construction of mathematical models that describe the tunable transport of solitons in a given structure.
Since most losses in networks occur at the branching points (vertices), ensuring freedom from reflection or absorption at these points is a key problem in avoiding signal loss. Solving such problems requires effective mathematical models that describe the reflection-free propagation of solitons in networks and branched structures.

A powerful mathematical tool for solving the problem of reflectionless soliton propagation is the imposition of so-called \textit{transparent boundary conditions (TBCs)} on a wave equation describing soliton transport. 
When projecting the problem onto networks, the transparent boundary conditions should be imposed at the network vertices (nodes). 
The problem of optimising signal transfer in opto-electronic networks thus leads to the development of models for nonlinear networks with transparent nodes. 
Apart from opto-electronic networks, the Manakov system in branched domains finds application in modelling the dynamics of vector solitons in BEC on branched traps, in the realization of soliton-based logic gates in networks and in the transport of vector solitons in branched thin crystals.

In contrast to linear evolution equations, the problem of designing TBCs for nonlinear equations cannot be solved by simple factorization of the differential operator or by direct use of pseudodiffential operators. 
However, for special cases of the nonlinear Schr\"odinger equation it is possible to formulate the exact TBCs in closed form by using the so-called \textit{potential approach} \cite{Zheng2006, Antoine}.
Such an approach has recently been successfully applied to NLSE on metric graphs. 
 Later, the approach was used for modelling the reflectionless propagation of the Manakov soliton on a line \cite{Jambul3}.
 In \cite{TBCSGE} the potential approach was extended to the sine-Gordon equation on a line describing the reflectionless kink propagation in 1D space.

 Here, we extend this promising concept to the Manakov system on networks by modeling the latter as \textit{metric graphs}. 
Metric graphs are domains consisting of 1D wires connected at the nodes. 
The connection rule is called the topology of a graph and is given for any graph by the so-called adjacency matrix \cite{Jambul1, Jambul}. 
We note that nonlinear evolution equations on branched domains and networks have attracted much attention in the literature in the last decade (see Refs.~\cite{Zarif}--\cite{Mashrab2022} and recent review paper \cite{DP2022}). 

This paper is organized as follows. 
In the next section, we briefly recall soliton solutions and conserving quantities for the Manakov system on a line. 
Section~III contains a brief discussion of transparent boundary conditions for the Manakov system on a line. 
In Section~IV, we formulate the problem of TBCs for the Manakov system on metric graphs together with some analytical and numerical results. 
Finally, Section~V contains the concluding remarks.

\section{Transparent boundary conditions for the Manakov system on a line}

\subsection{Soliton solutions of the Manakov system}
The \textit{Manakov system} is a two-component coupled nonlinear Schr\"odinger equation, which is explicitly given as 
\begin{equation}\label{ms1}    
\begin{split}
  \iu\partial_t\Psi_1+\frac{1}{2}\partial_x^2\Psi_1+(|\Psi_1|^2+|\Psi_2|^2)\Psi_1&=0,\\
  \iu\partial_t\Psi_2+\frac{1}{2}\partial_x^2\Psi_2+(|\Psi_1|^2+|\Psi_2|^2)\Psi_2&=0,     
\end{split}
\end{equation}
where $(\Psi_1,\Psi_2)=\bigl(\Psi_1(x,t),\Psi_2(x,t)\bigr)$, $x\in\mathbb{R}$, $t>0$.
It was introduced first by Manakov \cite{Manakov} to describe statio\-nary self-focusing electromagnetic waves in homogeneous waveguide channels.  
The \textit{single-soliton solution} of the Manakov system can be written as 
\begin{equation}
  (\Psi_1^*, \Psi_2^*) = \iu\Bigl( \frac{c}{|c|}\Bigr)
  \frac{\eta \exp|2\iu(\eta^2-\xi^2)t -2\iu x\xi|}{\cosh[2\eta(x+x_0 +2\xi t)]},
\end{equation}
where $x_0 =\ln (2\eta/|c|)/2\eta$ is the initial position of the soliton and
 $\bigl( c \equiv (c_{11},c_{21})]\bigr)$ is the unit vector that determines the soliton polarization. 
 The parameters $\xi$ and $\eta$ denote the speed and amplitude of the soliton, respectively.

Multi-soliton solutions of the Manakov system can be obtained by Hirota's bilinearization method \cite{Lakshmanan5, Lakshmanan6}.
Eq.~\eqref{ms1} admits infinitely many preserving quantities, which implies their integrability. Physically, the two most important conserving quantities, the norm $N$ and the energy $E$, determined as in \cite{ismail20081}, are respectively
\begin{equation*}
    N=\int_{-\infty}^{\infty} \bigl(|\Psi_1|^2 +|\Psi_2|^2\bigr)\,dx
\end{equation*}

\begin{equation}\label{eq:ener}
    E=\int\limits_{-\infty}^{+\infty}{\biggl(\sum_{m=1}^{2}{\frac{1}{2}\Bigl|\frac{\partial\Psi_m}{\partial x}\Bigr|^2}
    -\frac{1}{2}\sum_{m=1}^{2}{\left|\Psi_m\right|^4}-\left|\Psi_1\right|^2\left|\Psi_2\right|^2\biggr)dx}.
\end{equation}
In \cite{Sensei} the norm and energy conservation laws have been used to derive vertex boundary conditions for the Manakov system on a metric graph.

\subsection{Transparent boundary conditions}
Here we briefly recall, following Ref.~\cite{Jambul3}, the problem of transparent boundary conditions for the Manakov system.  
In the framework of the potential approach, the Manakov system~\eqref{ms1} can be formally written as a system of linear PDEs:
\begin{equation}\label{ms2}    
\begin{split}
   \iu\partial_t\Psi_1+\frac{1}{2}\partial_x^2\Psi_1+V(x,t)\Psi_1&=0,\\
   \iu\partial_t\Psi_2+\frac{1}{2}\partial_x^2\Psi_2+V(x,t)\Psi_2&=0,
\end{split}
\end{equation}
where the potential $V(x,t)$ is given as  
\begin{equation*}
    V(x,t)=|\Psi_1|^2+|\Psi_2|^2.
\end{equation*}
Introducing  the new vector function as \cite{Jambul3}
\begin{equation} \label{nvf1}
    \begin{split}
  v(x,t)&=e^{-\iu\nu(x,t)}\Psi(x,t),\\
  \Psi(x,t)&=\begin{pmatrix}\Psi_1(x,t)\\\Psi_2(x,t)\end{pmatrix},\quad
  v(x,t)=\begin{pmatrix}v_1(x,t)\\v_2(x,t)\end{pmatrix},
\end{split}
\end{equation}
where
\begin{equation}  \label{eq:nu}
    \begin{split}
    \nu(x,t)&=\int_0^tV(x,\tau)\,d\tau\\
    &=\int_0^t\bigl(|\Psi_1(x,\tau)|^2+|\Psi_2(x,\tau)|^2\bigr)\,d\tau.
\end{split}
\end{equation}
from Eq.~\eqref{ms2}, one can get \cite{Jambul3}
\begin{equation}\label{ms3}
    L(x,t,\partial_x,\partial_t)v
    =\iu\partial_tv
    +\frac{1}{2}\partial_x^2v+A\partial_xv+Bv=0,
\end{equation}
where
$A=\iu\partial_x\nu$, $B=\frac{1}{2}\bigl(\iu\partial_x^2\nu-(\partial_x\nu)^2\bigr)$.
Using pseudo-differential operators symbolic calculations we can write the operator $L$ as \cite{Jambul3}
\begin{multline}\label{ms4}
    L=\Bigl(\frac{1}{\sqrt{2}}\partial_x+\iu\Lambda^-\Bigr)\Bigl(\frac{1}{\sqrt{2}}\partial_x+i\Lambda^+\Bigr)\\
    =\frac{1}{2}\partial_x^2+\frac{\iu}{\sqrt{2}}(\Lambda^++\Lambda^-)\partial_x
    +\frac{\iu}{\sqrt{2}}\Op(\partial_x\lambda^+)-\Lambda^-\Lambda^+,
\end{multline}
where $\lambda^+$ is the principal symbol of the operator $\Lambda^+$ and $\Op(p)$ denotes the associated operator of a symbol $p$. 
The Eqs.~\eqref{ms3} and \eqref{ms4} 
provide the transparent boundary conditions for the Manakov system \eqref{ms1} in the form of the following symbolic system of equations \cite{Jambul3}:
\begin{align}
  \frac{\iu}{\sqrt{2}}(\lambda^++\lambda^-)&=a,\nonumber\\
  \frac{\iu}{\sqrt{2}}\partial_x\lambda^+-\sum_{\alpha=0}^{+\infty}\frac{(-\iu)^\alpha}{\alpha!}
  \partial_\tau^\alpha\lambda^-\partial_t^\alpha\lambda^+&=-\tau+b,\label{sos1}
\end{align}
where $\Op(a)=A$ and $\Op(b)=B$ can be set as $a=A$ and $b=B$ due to the fact that these two functions correspond to zero order operators. 
An asymptotic expansion in the inhomogeneous symbols
\begin{equation}
    \lambda^\pm\sim\sum_{j=0}^{+\infty}\lambda_{1/2-j/2}^\pm,\label{expl}
\end{equation}
allows to write the TBCs in different orders of approximation (see Ref.~\cite{Jambul3} for details). For the first order approximation holds
\begin{eqnarray}
  \frac{1}{\sqrt{2}}\partial_x\Psi_1|_{x=0}-e^{-\iu\frac{\pi}{4}}e^{\iu\nu}\cdot\partial_t^{1/2}(e^{-\iu\nu}\Psi_1)|_{x=0}=0,\nonumber\\
  \frac{1}{\sqrt{2}}\partial_x\Psi_2|_{x=0}-e^{-\iu\frac{\pi}{4}}e^{\iu\nu}\cdot\partial_t^{1/2}(e^{-\iu\nu}\Psi_2)|_{x=0}=0,\label{foal}\\
  \frac{1}{\sqrt{2}}\partial_x\Psi_1|_{x=L}+e^{-\iu\frac{\pi}{4}}e^{\iu\nu}\cdot\partial_t^{1/2}(e^{-\iu\nu}\Psi_1)|_{x=L}=0,\nonumber\\
  \frac{1}{\sqrt{2}}\partial_x\Psi_2|_{x=L}+e^{-\iu\frac{\pi}{4}}e^{\iu\nu}\cdot\partial_t^{1/2}(e^{-\iu\nu}\Psi_2)|_{x=L}=0.\label{foar}
\end{eqnarray}
We note that unlike the standard Dirichlet, Neumann or Robin boundary conditions, the boundary conditions given by Eqs.~\eqref{foar}, are quite complicated and can be implemented only numerically. 

\begin{figure}[th!]
\includegraphics[width=0.45\textwidth]{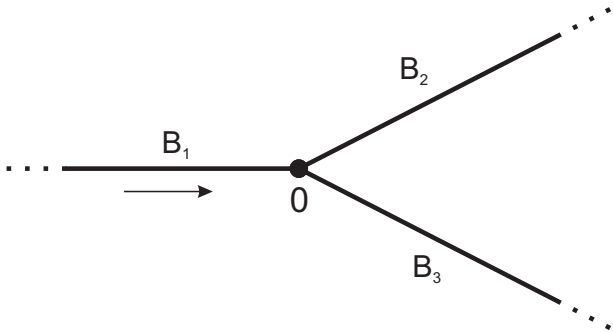}
\caption{Sketch of a star graph with 3 semi-inﬁnite bonds.}
\label{fig:1}
\end{figure}

\section{Transparent boundary conditions for the Manakov system on metric graphs}

\subsection{Manakov system on a metric star graph}
In this section we consider the dynamics of Manakov solitons on a metric star graph (see Fig.~\ref{fig:1}), focusing on the reflectionless transmission of solitons through the vertex boundary. 
The Manakov system on graphs has already been considered in \cite{Sensei}, where the integrability of the problem under certain conditions was shown and explicit soliton solutions were derived. 
From the point of view of formulation and solution of the evolution equations, a metric graph is a domain consisting of branched wires connected at the nodes. 
On each bond (arm) of the graph, the dynamics of the vector solitons is determined by a set of integrable two-component coupled nonlinear Schr\"odinger equations of Manakov type given by \cite{Sensei}
\begin{subequations}
\begin{align}    
   &\iu\partial_t u_j+\frac{1}{2}\partial_x^2 u_j+\beta_j(|u_j|^2+|v_j|^2) u_j=0,\label{mss1}\\
   &\iu\partial_t v_j+\frac{1}{2}\partial_x^2 v_j+\beta_j(|u_j|^2+|v_j|^2) v_j=0,\label{mss2}
\end{align}
\end{subequations}
where $u_j, v_j, j=1,2,3$, are the components of the Manakov soliton, 
$j$ represents the bond index and $\beta_j$ is the nonlinearity coefficient at bond $j$. 

We denote $\Psi_j=(u_j,v_j)^\top$ and set the boundary conditions for the vertex resulting from the weight continuity
\begin{equation}\label{vbc1}
    \alpha_1\Psi_1(0,t)=\alpha_2\Psi_2(0,t)=\alpha_3\Psi_3(0,t) 
\end{equation}
and the generalized Kirchhoff rules
\begin{equation}\label{vbc2}
   \frac{1}{\alpha_1}\frac{\partial\Psi_1}{\partial x}\Bigr|_{x=0}
   =\frac{1}{\alpha_2}\frac{\partial\Psi_2}{\partial x}\Bigr|_{x=0}+\frac{1}{\alpha_3}\frac{\partial\Psi_3}{\partial x}\Bigr|_{x=0},
\end{equation}
where $\alpha_j (j=1,2,3)$ are nonzero real constants.

Without loss of generality of the approach, we can assume that $\alpha_j=\sqrt{\beta_j}$. Then we assume that the following condition is satisfied
\begin{equation} 
    \frac{1}{\beta_1} = \frac{1}{\beta_2} +\frac{1}{\beta_3}.
\end{equation}
The soliton solutions of the Manakov system on metric star graph can be written as \cite{Sensei}
\begin{equation}
    \Psi_j(x) =\frac{1}{\sqrt{\beta_j}}\Phi_j (x),
\end{equation}
where $\Phi_j (x)$ are the soliton solutions of the Manakov system \eqref{ms1} on a line.

\subsection{TBC for the Manakov system on a metric star graph}
In this section, we derive the TBCs for the Manakov system on graphs and prove that they are equivalent to the vertex boundary conditions given by Eqs.~\eqref{vbc1} and \eqref{vbc2}, provided that the sum rule in Eqs.~\eqref{sumrule1} is satisfied.
The system \eqref{mss1}-\eqref{mss2} can be formally reduced to a system of linear PDEs and written as follows 
\begin{equation}\label{msf}    
    \iu\partial_t\Psi_j+\frac{1}{2}\partial_x^2\Psi_j+V_j(x,t)\Psi_j=0,
\end{equation}
where $\Psi_j=(u_j,v_j)^\top$ and the potential, $V_j(x,t)$ is given as  
\begin{equation*}
   V_j(x,t)=\beta_j(|u_j|^2+|v_j|^2).
\end{equation*}

To impose TBCs for Eq.~\eqref{msf}, we split the whole domain (graph) into two domains, which we call ``interior'' ($-\infty<x<0$) and ``exterior'' ($0 < x < \infty$). 
These terminologies were borrowed from the original papers \cite{Arnold1998, Ehrhardt1999, Ehrhardt2001}, in which the basic idea of constructing TBCs was proposed. 
Accordingly, we have interior and exterior problems. 
The interior problem for $B_1$ can be written as
\begin{align}    
  &\iu\partial_t\Psi_1+\frac{1}{2}\partial_x^2\Psi_1+V_1(x,t)\Psi_1=0,\label{intprob}\\
  &\Psi_1\bigr|_{t=0}=\psi^I(x),\nonumber\\
  &\partial_x\Psi_1\bigr|_{x=0}=(T_0\Psi_1)\bigr|_{x=0}.\nonumber
\end{align}

The exterior problems for $B_{2,3}$ are given as
\begin{align}    
   &\iu\partial_t\Psi_{2,3}+\frac{1}{2}\partial_x^2\Psi_{2,3}+V_{2,3}(x,t)\Psi_{2,3}=0,\label{extprob}\\
   &\Psi_{2,3}\bigr|_{t=0}=0,\nonumber\\
   &\partial_x\Psi_{2,3}\bigr|_{x=0}=\Phi_{2,3}(t), \ \ \Phi_{2,3}(0)=0,\nonumber\\
   &(T_0\Phi_{2,3})\bigr|_{x=0}=\partial_x\Psi_{2,3}\bigr|_{x=0}.\nonumber
\end{align}

We introduce the following new vector function as
\begin{equation}\label{eq:vf1}
\begin{split}
  w(x,t)&=e^{-\iu\nu_{2,3}(x,t)}\Psi_{2,3}(x,t),\\
  \Psi_{2,3}(x,t)&=\begin{pmatrix}u_{2,3}(x,t)\\ v_{2,3}(x,t)\end{pmatrix},\quad w(x,t)=\begin{pmatrix}\omega_1(x,t)\\ \omega_2(x,t)\end{pmatrix},
\end{split}
\end{equation}
where
\begin{equation}\label{eq:nu2}
    \begin{split}
    \nu_{2,3}(x,t)&=\int_0^tV_{2,3}(x,\tau)\,d\tau\\
    &=\beta_{2,3}\int_0^t\bigl(|u_{2,3}(x,\tau)|^2+|v_{2,3}(x,\tau)|^2\bigr)\,d\tau.
\end{split}
\end{equation}

Using the results of the recent work \cite{Jambul3}, the formal TBCs for $\Psi_{2,3}$ at $x=0$ can be written as 
\begin{multline*}
    \partial_x\Psi_{2,3}|_{x=0}=\sqrt{2}e^{-\iu\frac{\pi}{4}}e^{\iu\nu_{2,3}}\cdot\partial_t^{1/2}(e^{-\iu\nu_{2,3}}\Psi_{2,3})|_{x=0}\\
     +\iu\frac{1}{4}\partial_xV_{2,3}e^{\iu\nu_{2,3}}I_t(e^{-\iu\nu_{2,3}}\Psi_{2,3})|_{x=0}=0,
\end{multline*}
where the fractional 1/2 derivative is given by
\begin{equation*}
    \partial^{1/2}_t f(t)=\frac{1}{\sqrt{\pi}}\partial_t\int_0^t{\frac{f(\tau)}{\sqrt{t-\tau}}d\tau},
\end{equation*}
and 
\begin{equation*}
   I_t(f)=\int_0^t{f(\tau)\,d\tau}.
\end{equation*}

Using the vertex boundary conditions \eqref{vbc1} and \eqref{vbc2} we have
\begin{equation} 
    \nu_{2,3}(0,t)=\nu_1(0,t),    
\end{equation}
\begin{equation}
      \frac{1}{\beta_1}\partial_x V_1|_{x=0}=\frac{1}{\beta_2}\partial_x V_2|_{x=0}+\frac{1}{\beta_3}\partial_x V_3|_{x=0},
\end{equation}
which leads to
\begin{multline*}
    \partial_x\Psi_{2,3}|_{x=0}=\sqrt{2}\sqrt{\frac{\beta_1}{\beta_{2,3}}}e^{-\iu\frac{\pi}{4}}e^{\iu\nu_1}\cdot\partial_t^{1/2}(e^{-\iu\nu_1}\Psi_1)|_{x=0}\\
     +\iu\frac{1}{4}\sqrt{\frac{\beta_1}{\beta_{2,3}}}\partial_xV_{2,3}e^{\iu\nu_1}I_t(e^{-\iu\nu_1}\Psi_1)|_{x=0}=0,
\end{multline*}

Now, from the boundary conditions \eqref{vbc1} and \eqref{vbc2} we obtain
\begin{align}
   \partial_x\Psi_1\bigr|_{x=0}
   &=\frac{\sqrt{\beta_1}}{\sqrt{\beta_2}}\partial_x \Psi_2\bigr|_{x=0}+\frac{\sqrt{\beta_1}}{\sqrt{\beta_3}}\partial_x \Psi_3\bigr|_{x=0}\nonumber\\
    =\beta_1&\Bigl(\frac{1}{\beta_2}+\frac{1}{\beta_3}\Bigr)\Bigr[\sqrt{2}e^{-\iu\frac{\pi}{4}}e^{\iu\nu_1}\cdot\partial_t^{1/2}(e^{-\iu\nu_1}\Psi_1)\Bigl]\big|_{x=0}\nonumber\\
    &+\iu\frac{1}{4}\partial_xV_1e^{\iu\nu_1}I_t(e^{-\iu\nu_1}\Psi_1)|_{x=0}=0.\label{tbc}
\end{align}

The boundary condition given by \eqref{tbc} coincides with that in Eq.~\eqref{foal} and thus provides reflection-free transmission for the
bond $B_1$ if the following sum rule is fulfilled:
\begin{equation}\label{sumrule1}  
     \beta_1\left(\frac{1}{\beta_2}+\frac{1}{\beta_3}\right)=1.  
\end{equation}
Thus, the satisfaction of the sum rule \eqref{sumrule1} implies
that the vertex boundary conditions \eqref{vbc1} and \eqref{vbc2} become equivalent to the TBCs at the vertex of the graph.

\begin{figure}[th!]
\includegraphics[width=0.45\textwidth]{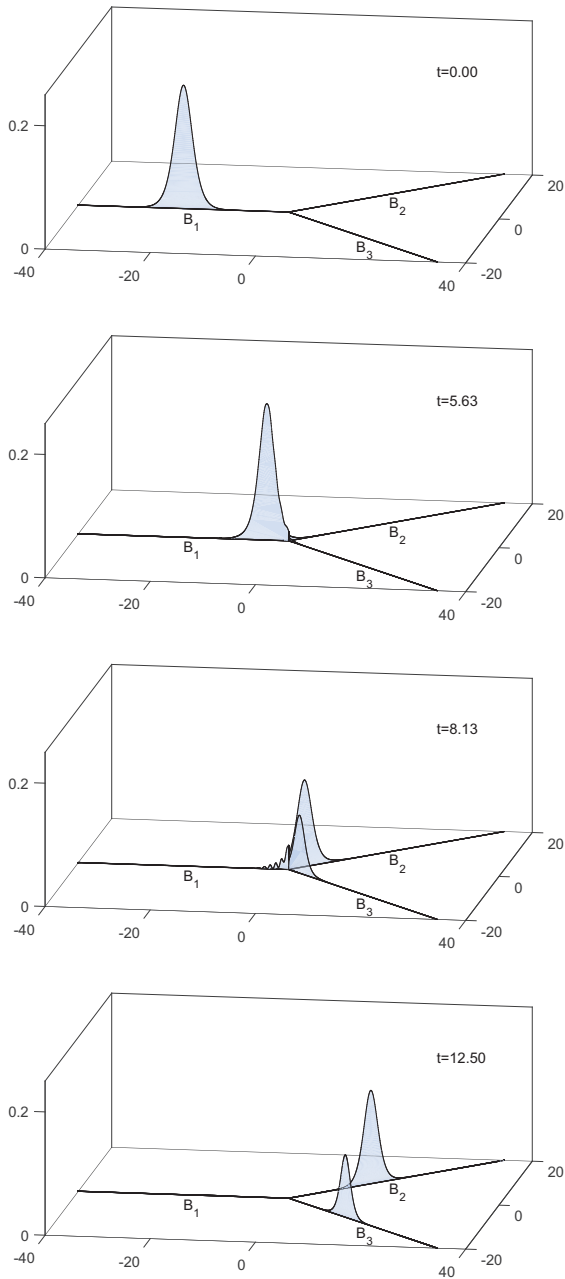}
\caption{Soliton dynamics plotted at different time instances for the regime when the sum rule is fulfilled (no reflection occurred): $\beta_1=1.2$, $\beta_2=2$, $\beta_3=3$.}
\label{fig:2}
\end{figure}

\section{Numerical Experiment}
In this section, we show numerically that satisfying the condition \eqref{sumrule1} allows reflectionless transmission of a soliton through the vertex of a graph with boundary conditions in the form of \eqref{vbc1} and \eqref{vbc2}. 
The experimental setup consists of a star-shaped graph with three bonds (see Fig.~\ref{fig:1}). 
We consider the soliton going from the first bond to the second and third, i.e.,
the initial condition is compactly supported in the first bond. 
As initial condition we choose a single soliton from the exact solution given by
\begin{equation*}
    G(x)=\sqrt{\sigma}\sech\bigl[\sqrt{2\sigma}(x-x_0)\bigr]\exp\bigl[\iu\sqrt{2}p(x-x_0)\bigr],
\end{equation*}
so that
\begin{align*}
  &\Psi_1(x,0)=G(x)\begin{pmatrix}1\\1\end{pmatrix},\\   
  &\Psi_2(x,0)=\begin{pmatrix}0, 0\end{pmatrix}^\top,\quad
  \Psi_3(x,0)=\begin{pmatrix}0, 0\end{pmatrix}^\top.
\end{align*}

\begin{figure}[th!]
\includegraphics[width=0.45\textwidth]{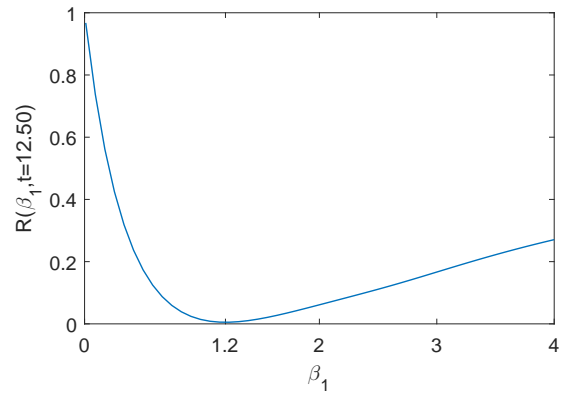}
\caption{Dependence of the vertex reflection coefficient $R$ on the
parameter $\beta_1$ when time elapses ($t=12.50$). 
For fixed $\beta_2=2$ and $\beta_3=3$, $R=0$ when $\beta_1=1.2$.}
\label{fig:3}
\end{figure}

In our experiments, we used the explicit midpoint rule \cite{frutos1992}, the
so-called leapfrog ﬁnite-difference method, which was also used in our recent work \cite{Jambul3}. 
We chose the following system parameters: 
The lengths of the bonds $B_1, B_2, B_3$ are $L_1 = L_2 = L_3 = 40$, accordingly the parameters of the initial condition are $\sigma=1$, $p=1$, $x_0=-20$ and the space discretization $\Delta x=0.05$, the time step $\Delta t=0.00125$. 
The evolution of the right-travelling single soliton is shown in Fig.~\ref{fig:2} in four successive time steps. 
From this plot, the reflectionless transmission of the soliton is evident.

Finally, in order to show that in the case where the sum rule is violated
the transmission of the soliton is accompanied by reflections, we have plotted in
Fig.~\ref{fig:3} the reflection coefficient, which is determined as the ratio of the partial norm for the first bond to the total norm, i.e.
\begin{equation*}
  R=\frac{N_1}{N_1+N_2+N_3},
\end{equation*}
as a function of $\beta_1$ for the ﬁxed values of $\beta_2$ and $\beta_3$. 
From this plot it can be seen that the reflection coefficient becomes zero for the value of $\beta_1$ that satisfies the sum rule \eqref{sumrule1}.

\section{Conclusions}
In this paper, by extending our previous study presented in Ref.~\cite{Jambul3}, we have studied the problem of transparent Manakov networks with transparent nodes. 
The latter implies the absence of backscattering at the nodes.
The concept of transparent boundary conditions is applied to the Manakov system on metric graphs to model the reflectionless transmission of vector solitons through the graph vertex.
It is shown that for the case where the nonlinearity coefficients satisfy the sum rule in Eq.~\eqref{sumrule1}, the vertex boundary conditions in terms of weight continuity and Kirchhoff rules become equivalent to the transparent BCs. Although the above study is focused on star branched graph only, the approach we proposed can be directly applied for wide class of graph topologies, such as e.g., tree graph or H-graph, etc. The only restriction for graph architecture is that it should have one incoming and at least two outgoing semi-infinite bonds and arbitrary graph in between these bonds. The results obtained in this paper have direct application in modeling reflectionless (lossless) signal propagation in opto-electronic networks, optical waveguides, tunable transport of BEC in branched traps, etc.


\end{document}